\documentclass[aps,pre,twocolumn,letter,longbibliography,superscriptaddress]{revtex4-2}
\usepackage[english]{babel}

\usepackage{graphicx}
\DeclareGraphicsExtensions{.pdf}

\usepackage{float}
\usepackage{amsmath}
\usepackage{color}
\usepackage{ulem}
\usepackage{hyperref}

\newcommand{\be}{\begin{equation}}
\newcommand{\ee}{\end{equation}}

\definecolor{green}{rgb}{0.1, 0.8, 0.1}

\begin{document}
\title{Boundary-Mediated Phases of Self-Propelled Kuramoto Particles}
\author{Francesco Arceri}
\affiliation{INFN, Sezione di Padova, Via Marzolo 8, I-35131 Padova, Italy}
\author{Vittoria Sposini}
\affiliation{Department of Physics and Astronomy, University of Padova, Via Marzolo 8, I-35131 Padova, Italy}
\author{Enzo Orlandini}
\affiliation{Department of Physics and Astronomy, University of Padova, Via Marzolo 8, I-35131 Padova, Italy}
\author{Fulvio Baldovin}
\affiliation{Department of Physics and Astronomy, University of Padova, Via Marzolo 8, I-35131 Padova, Italy}
\date{\today}

\begin{abstract}
Active agents can transfer energy to their environment through collective motion, generating accumulation patterns near confining obstacles. Here we investigate how the nature of the microscopic drive—self-propulsion or velocity alignment—selects distinct accumulation patterns, leading to either delocalized or compact clustered states. We first characterize the dynamical regimes emerging from the interplay of these two driving mechanisms under perfectly reflective or smooth boundary conditions. We then introduce boundary friction and observe a drastic change in the accumulation patterns, with new dynamical phases that are absent in the previous case. By connecting emergent macroscopic structures to their underlying microscopic interactions, this work provides a practical route to infer the dominant interaction ruling boundary-mediated collective behavior, with applications ranging from single-cell migration to bio-inspired robotics.
\end{abstract}

\maketitle

\section{Introduction}
From bacterial colonies in the intestinal lining~\cite{bricard_emergence_2013} to self-propelled robots in circular arenas~\cite{deseigne_vibrated_2012, weber_long-range_2013}, active agents spontaneously form dynamic structures in response to changes in the mechanical environment. The presence of confinement imposed by obstacles and walls triggers collective behaviors such as different accumulation states and the spontaneous emergence of coherent flows~\cite{caprini_dynamical_2024, ugolini_microfluidic_2024}. These effects are not only central to understanding how collective reconfigurations shape the response of living organisms across multiple scales, from microbial transport in porous media~\cite{de_anna_chemotaxis_2021} to responses to threats in large animal groups~\cite{deng_spontaneous_2021}; they also provide a theoretical basis for cytoskeletal remodeling in cell migration~\cite{zhang_drosophila_2011, feng_microtubule-binding_2017} and for harnessing bacteria as biological engines in micro-robotic applications~\cite{di_leonardo_bacterial_2010}.

The theoretical description of collective behavior in active matter typically relies on simplified models that capture the microscopic conversion and macroscopic redistribution of energy occurring on distinct length and time scales. A common mechanism is persistent self-propulsion, an internal force which drives motility at the level of individual agents; this is the case for certain bacteria able to migrate along nutrient gradients via chemotaxis~\cite{moore_physics_2024} or algae moving towards light gradients via phototaxis~\cite{tsang_light-dependent_2024}. Another mechanism underlying collective dynamics is velocity alignment, exemplified by the coordinated motion of bird flocks or insect swarms~\cite{cavagna_flocking_2015}, and by epithelial cells forming confluent monolayers~\cite{park_collective_2016, ilina_cellcell_2020}. The dynamics of such polar active matter arise from local interactions that synchronize the direction of motion among self-propelled agents. Coarse-grained models of aligning active matter, including the Vicsek and Kuramoto frameworks~\cite{vicsek_novel_1995-1, acebron_kuramoto_2005}, capture this collective behavior through simple coupling rules that promote orientational order. For what follows, we adopt the Kuramoto formulation, where synchronization is modeled by a reciprocal velocity coupling between agents, which provides a simple description of alignment-driven collective motion in active systems~\cite{chepizhko_relation_2010, martin-gomez_collective_2018}.

While self-propulsion and polar alignment are well established features of active matter, the phenomenology arising from their simultaneous action under confinement remains much less explored. For self-propelled particles without any explicit alignment interactions, a number of studies have focused on dense regimes where excluded-volume interactions compete with activity and lead to motility-induced phase separation (MIPS) between a dense and a dilute phase~\cite{buttinoni_dynamical_2013-2, cates_motility-induced_2015}, the former more prone to accumulate near surfaces than the latter. For alignment-dominated systems, research has instead concentrated on bulk flocking phenomena and the emergence of long-range order~\cite{baglietto_nature_2009, cavagna_flocking_2015}. Confinement itself is known to induce clustering and ordering in certain colloidal suspensions and granular systems~\cite{solano-cabrera_self-assembly_2025}, yet it remains unclear how different driving forces interact in such settings, what types of boundary accumulation patterns emerge, and what is the role of boundary friction.

In this work, we show how boundary accumulation patterns depend on the relative strength of self-propulsion and polar alignment. We first study self-propelled Kuramoto particles (SPKPs) confined by a frictionless circular boundary, where the two driving forces are independently tuned~\cite{martin-gomez_collective_2018, liebchen_chiral_2022}. By systematically varying the time scales associated with self-propulsion reorientation and alignment relaxation, we identify distinct clustering modes near boundaries, ranging from annular to compact droplet-like clusters. We then include boundary friction by confining particles within a ring polymer and observe how these accumulation modes are altered by the presence of roughness. Notably, we find that friction counteracts alignment-mediated cohesion, suppressing delocalized accumulation structures so that only thicker clusters that partially span the boundary may survive. We also identify a new phase where particles accumulate near the boundary with no net collective motion. Our results elucidate the interplay between mechanical confinement and emerging collective behavior, laying the groundwork for future studies on harnessing collective motion in living and artificial confined active systems.

\section{Numerical methods}
We study a two-dimensional system of SPKPs confined within a circular wall of radius $R$. Particle diameters are drawn from a log-normal distribution with 20$\%$ polydispersity to suppress boundary-induced ordering. The average particle diameter, $\sigma$, is taken as the unit of length, and all particle masses are set to $m = 1$. We fix the number of particles to $N=1024$ and limit the study to a very dilute system of area fraction $\sum_i (\sigma_i / 2)^2 / R^2 = 0.014$. The chosen area fraction lies within the dilute regime observed in suspensions of microbial communities~\cite{brun-cosme-bruny_effective_2019, ghosh_cross_2022} and active colloids~\cite{morin_distortion_2017, modica_porous_2022}. Particles are evolved via driven Brownian dynamics
\begin{equation}\label{eq1}
\frac{d\vec{r}_i}{dt} = -\mu \sum_{j\neq i}\vec{\nabla}U_{ij}^{WCA} \, + \, v_0 \hat{n}_i \, + \, \mu \vec{F}^w_i \, ,
\end{equation}
where steric interactions are modeled by the purely repulsive Weeks–Chandler–Andersen (WCA) potential
\begin{equation}
U^{WCA}_{ij} =
\begin{cases}
4\varepsilon \left[ 
\left( \dfrac{\sigma_{ij}}{r_{ij}} \right)^{12}
-
\left( \dfrac{\sigma_{ij}}{r_{ij}} \right)^{6}
\right]
+ \varepsilon,
& r_{ij} \leq 2^{1/6}\sigma_{ij}, \\[10pt]
0, & r_{ij} > 2^{1/6}\sigma_{ij}.
\end{cases}
\end{equation}
where $r_{ij}$ is the distance between particle $i$ and particle $j$ and $\sigma_{ij} = (\sigma_i + \sigma_j)/2$ is the average particle diameter. The other terms in the equation describe the active driving, $v_0\hat{n}_i$, and the interaction with the boundary, $\mu \vec{F}^w_i$. We set the energy scale of the potential to $\varepsilon=1$ and define $\sigma \sqrt{m / \varepsilon}$ as the unit of time, $\tau_i$. The mobility is set to $\mu = 10^{-2}\tau_i / m$ such that the inertial relaxation time $m \mu$ is negligible. The magnitude of the active velocity $v_0$ is fixed to $2 \sigma / \tau_i$ corresponding to a kinetic energy scale $m v_0^2/2 = 2 \varepsilon$, comparable to the interaction energy scale $\varepsilon$. Particle velocity components are initialized from a Gaussian distribution with zero mean and variance $v_0^2/2$. The orientation of the active force acting on particle $i$ is defined by a unit vector $\hat{n}_i(t) = (\cos\theta_i(t), \sin\theta_i(t))$, where the angle $\theta_i(t)$ evolves following the equation
\begin{equation}\label{eq2}
\frac{d\theta_i(t)}{dt} = \sqrt{\frac{2}{\tau_p}} \; \xi_i(t) \, - \, \frac{1}{\tau_K} \sum_{\substack{j \\ r_{ij} < R_K}} \sin(\varphi_{v,i}(t) - \varphi_{v,j}(t)) \, .
\end{equation}
Here, the first term on the r.h.s. of the equation is a scalar white noise, where $\tau_p$ is the persistence time, i.e. the typical time needed to change the direction of self-propulsion, and $\xi_i(t)$ is a Gaussian random process of mean zero and correlation $\langle \xi_i(t)\xi_i(t') \rangle = \delta(t-t')$. The second term is the Kuramoto interaction, where the timescale $\tau_K = \pi R_K^2 / K$ is the alignment time, with $K$ being the coupling strength and $R_K$ the interaction radius. In the following, we will vary $\tau_K$ by varying the coupling strength $K$, while keeping the interaction radius equal to $1.5\sigma$. Particles align by minimizing the sine of the difference between their velocity orientations, namely $\varphi_{v,i} = \arctan(v_{i,y}/v_{i,x})$. In the absence of the Kuramoto term, the model reduces to a system of Active Brownian particles which may acquire local velocity alignment from persistent collisions in the MIPS regime~\cite{digregorio_full_2018-1, caprini_spontaneous_2020}. However, this phenomenon is observed in much denser systems than those studied here, so we can attribute alignment solely to the Kuramoto term. Eqs. (\ref{eq1}) and (\ref{eq2}) are numerically integrated using the Euler-Maruyama scheme with a time step $\Delta t = 5 \times 10^{-5}$.

\begin{figure} [t]
\includegraphics[width=\columnwidth]{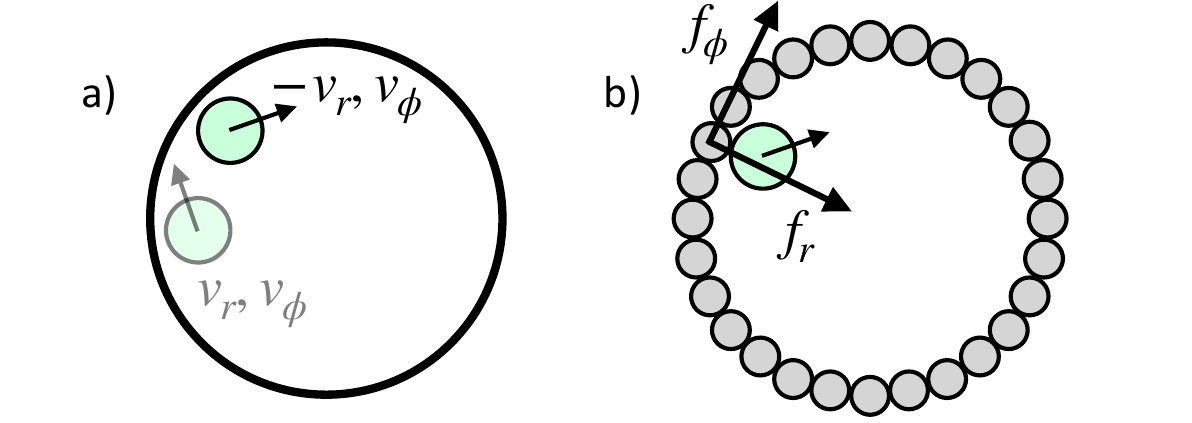}
\caption{(a) A particle before and after collision with a smooth boundary. (b) Particle-monomer interaction in a rough boundary with radial $f_r$ and tangential $f_\phi$ components.}
\label{fig:sketch} 
\end{figure}

By varying the particle-wall interaction $\vec{F}^w_i$ we model two types of boundaries schematically shown in Fig.~\ref{fig:sketch}: 1) \textit{smooth}, implemented as a perfectly reflective circle, where the radial velocity component $v_r$ of a particle colliding with the boundary is reversed upon collision while the tangential component $v_\theta$ remains unchanged; and 2) \textit{rough}, modeled as a ring polymer made of monomers that interact with particles via the WCA potential with energy scale $\varepsilon_w$. We set $\varepsilon_w = 10\varepsilon$ to model a significantly stiffer boundary compared to particle-particle interactions. More details are reported in Sec.~\ref{rough-boundary}.

\begin{figure*}
\centering
\includegraphics[width=1.9\columnwidth]{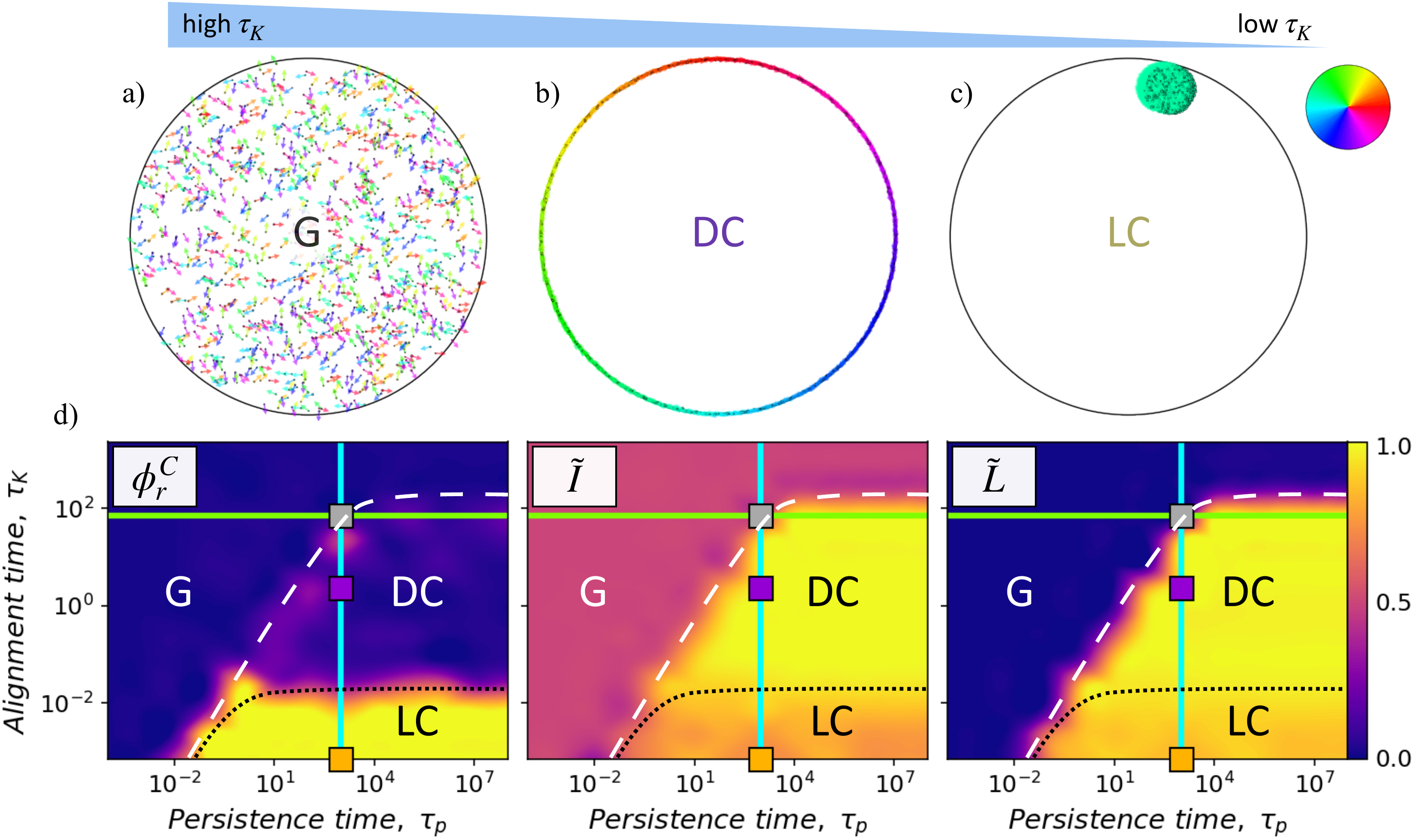}
\caption{Phases of SPKPs confined by a smooth boundary at fixed persistence time $\tau_p=10^3$. Values of $\tau_p$ and $\tau_K$ are reported in units of $\tau_i$. (a) Gas (G) phase at $\tau_K = 7.1\times10^1$. (b) Delocalized clustered (DC) phase at $\tau_K = 2.4$. (c) Localized clustered (LC) phase at $\tau_K = 7.1\times10^{-4}$. Particles and velocity arrows are color coded according to $\varphi_v$. (d) Diagrams in the plane $(\tau_p, \tau_K)$ color coded by $\phi_r^C$, $\tilde{I}$ and $\tilde{L}$. The vertical line at $\tau_p=10^3$ and the horizontal line at $\tau_K = 7.1\times10^1$ refer to the coordinate axes of Fig.~\ref{fig:moments}a. Dashed and dotted lines distinguish phases, while squares indicate the configurations shown above.}
\label{fig:smooth-phases} 
\end{figure*}

\section{Phases of confined Self-Propelled Kuramoto Particles}
We now examine how the interplay between self-propulsion and velocity alignment influences the collective dynamics of SPKPs confined by a smooth boundary. Particles move under a constant driving force and the higher their persistence, the longer they reside near the boundary before reorienting. Alignment interactions may then trigger the formation of groups of particles that slide along the boundary. As a result, clustering and collective motion spontaneously emerge. 

By varying $\tau_p$ and $\tau_K$, we identify three distinct phases reported in Fig.~\ref{fig:smooth-phases}. At short persistence time and weak alignment (large $\tau_K$), particles reorient more rapidly than they can align with their neighbors, leading to a homogeneous, weakly correlated gas of self-propelled particles. As $\tau_p$ increases at fixed $\tau_K$, particles travel longer along a fixed direction before reorienting. This enhances collisions between neighboring particles, promoting alignment and cluster formation. Once formed, clusters of aligned particles may reach the boundary and remain localized there due to their net persistent motion. Notably, for sufficiently small $\tau_K$, or sufficiently strong alignment, a delocalized clustered (DC) phase emerges where particles form a single annular cluster (roughly $2\sigma$ thick) that spans the whole circumference of the wall and rotates around its center. As alignment increases ($\tau_K$ decreases) while keeping $\tau_p$ fixed, particles are forced to sit at the interaction distance $R_K$ leading to the localized clustered (LC) phase where particles form one or more compact aggregates that slide along the boundary. Delocalized and localized clusters select a direction of rotation, either clockwise or counterclockwise, thus spontaneously breaking the symmetry of the confining geometry. We note that, when persistence is low ($\tau_p < 1$), the cluster direction of motion rapidly reorients and clusters are not bound to slide along the surface, see Video 3 in~\cite{supplemental}. We verified the occurrence of the different phases in a larger system composed of $N=32768$ particles and observed no significant variations.

To distinguish the observed dynamical patterns, we identify three key quantities whose behavior is summarized in Fig.~\ref{fig:smooth-phases}d. The first is the Kuramoto or \textit{localization parameter}~\cite{gil_common_2010},
\begin{equation}
\phi_r = \left| \frac{1}{N} \sum_{k=1}^{N} e^{i\varphi_{r,k}} \right| \, ,
\end{equation}
where $\varphi_{r,k}$ denotes the angular coordinate of particle $k$, defined by $\vec{r}_k= r_k(\cos\varphi_{r,k}, \sin\varphi_{r,k})$ and $i = \sqrt{-1}$. The localization parameter approaches zero when particles have a uniform distribution of angular coordinates (gas phase), and is equal to one if they share the same exact angular coordinate. Thus, $\phi_r$ quantifies whether particles in clusters are close to each other or not, being close to zero in the DC phase and close to one in the LC phase. However, $\phi_r$ fails to detect multiple localized clusters which we observe for $\tau_K > 2.4 \times 10^{-3}$, see Video 4 in~\cite{supplemental}, while a single cluster is always formed below this threshold. As an example, two clusters of equal size moving at an angular distance of $\pi$ would result in $\phi_r = 0$. To avoid this limitation, we compute the Kuramoto parameter for the largest cluster, $\phi_r^*$, identified using the DBSCAN algorithm~\cite{supplemental, hahsler_dbscan_2019}. When the largest cluster contains more than $20\%$ of the particles we use $\phi_r^*$, and $\phi_r$ otherwise; both are denoted by $\phi_r^C$.

\begin{figure*}
\centering
\includegraphics[width=1.9\columnwidth]{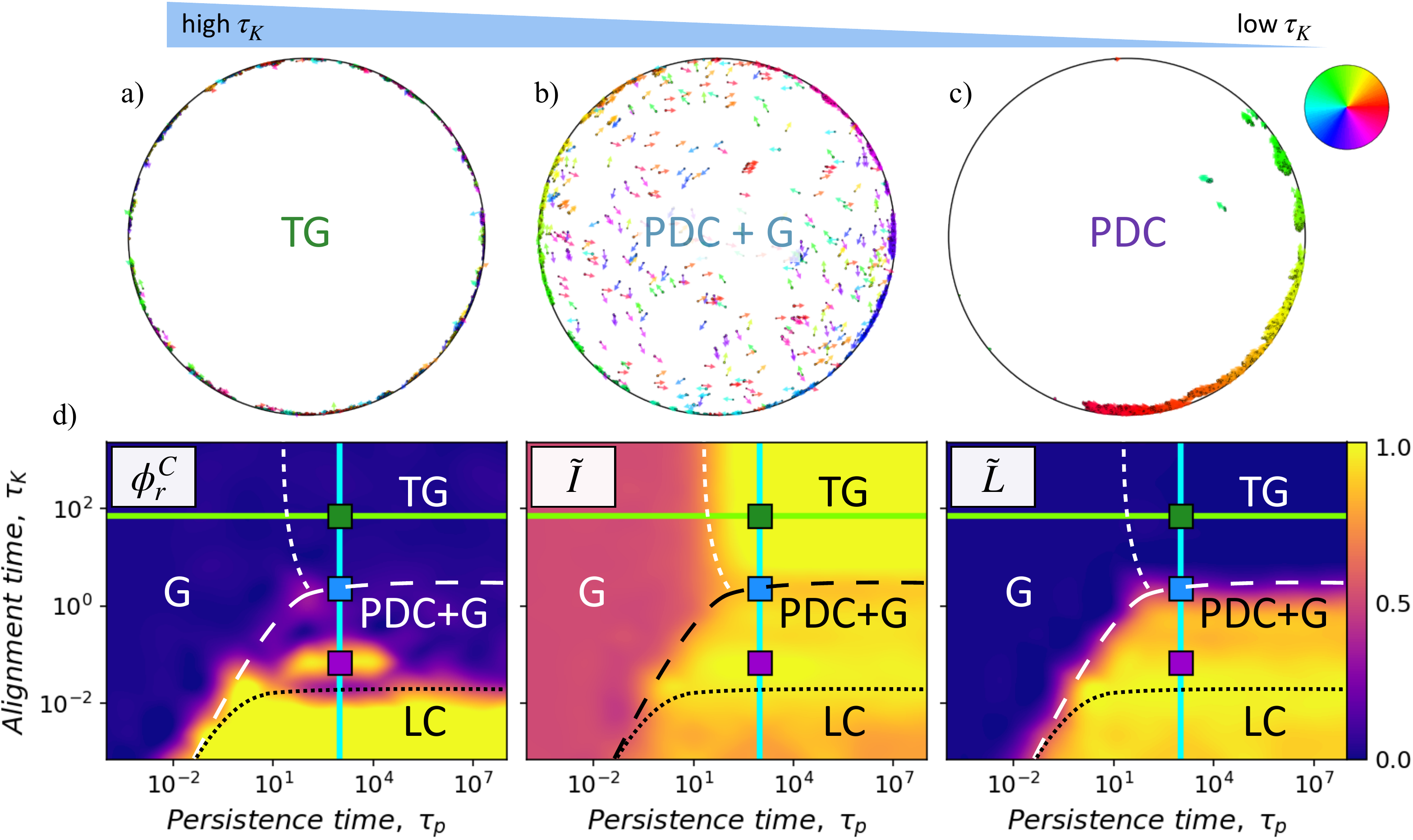}
\caption{Dynamical regimes of SPKPs confined by a rough boundary at fixed persistence time $\tau_p=10^3$. (a) Trapped gas (TG) phase at $\tau_K = 7.1\times10^1$. (b) Coexistence between partially delocalized clustered (PDC) phase and gas at $\tau_K = 2.4$. (c) PDC phase at $\tau_K = 7.1\times10^{-2}$. (d) Diagrams in the plane $(\tau_p, \tau_K)$ color coded by $\phi_r^C$, $\tilde{I}$ and $\tilde{L}$. The vertical line at $\tau_p=10^3$ and the horizontal line at $\tau_K = 7.1\times10^1$ refer to the coordinate axes of Fig.~\ref{fig:moments}a. Dashed and dotted lines distinguish phases, while squares indicate the configurations shown above.}
\label{fig:rough-phases} 
\end{figure*}

The second observable is the \textit{normalized moment of inertia},
\begin{equation}
\tilde{I} = \frac{m}{MR^2} \sum_{i=1}^N r_i^2 \, ,
\end{equation}
where $M R^2$ is the moment of inertia of a ring in which all the mass, $M = Nm$, is concentrated at a distance equal to the boundary radius $R$. This observable is $\tilde{I} \approx 0.5$ in the gas phase where the mass is uniformly distributed within the circle, and $\tilde{I} \approx 1$ in the DC phase where particles form a single ring-like cluster. The LC phase instead is characterized by intermediate values of $\tilde{I}$. Specifically, when particles form a single disk-like cluster of radius $R_c$ and mass $M$ that sits at a distance from the center equal to $R - R_c$, the normalized moment of inertia is
\begin{equation}\label{eq6}
\tilde{I} = \left(1 - \frac{R_c}{R} \right)^2 + \frac{1}{2}\left( \frac{R_c}{R}\right)^2 .
\end{equation}
This direct link between $\tilde{I}$ and the cluster radius provides an estimate of the area fraction, or density, within the cluster. For example, the configuration reported in Fig.~\ref{fig:smooth-phases}c corresponds to $\tilde{I} \approx 0.74$, and by inverting (\ref{eq6}) we get $R_c/R \approx 0.15$, i.e. a cluster radius of about $15\%$ of the wall radius. Thus, computing the number density as $n_c = N/\pi R_c^2$ yields an area fraction equal to $n_c a_p \approx 0.61$, where $a_p$ is the average particle area. Given the ratio between kinetic and potential energy scales, $m v_0^2 / 2\varepsilon = 2$, this density suggests that clusters are in a liquid-like state.

The third parameter we use to describe the phases of confined SPKPs is the \textit{normalized angular momentum},
\begin{equation}
\tilde{L} = \frac{m}{M v_0 R} \Bigg| \sum_{i=1}^N \vec{r}_i \times \vec{v}_i \Bigg| \, ,
\end{equation}
where $Mv_0R$ is the angular momentum if all the mass were concentrated at a distance $R$ from the center and rotated with angular velocity $\omega = v_0 / R$. We focus on the absolute value of the angular momentum to discard whether clusters rotate clockwise or counterclockwise. Figure~\ref{fig:smooth-phases}d clearly shows that $\tilde{I}$ and $\tilde{L}$ are strongly correlated in smooth boundary conditions. This is clear evidence that whenever particles are distributed close to the boundary, whether in ring-like or disk-like structures, they exhibit collective rotation.

\section{Influence of boundary roughness}\label{rough-boundary}
We now focus on rough boundary conditions and quantify the influence of roughness on the dynamical phases identified so far. To this end, we replace the reflective wall with an inextensible ring polymer composed of $N_m$ monomers of diameter $\sigma_m$ equal to half of the average particle diameter. The number of monomers is set to $N_m = 2 \pi R / \sigma_m$ (rounded to the nearest integer) and adjacent monomers are just in contact. Starting from configurations within smooth boundaries, we slightly rescale the radial particle coordinates by $0.01\%$ to prevent discontinuities in potential energy. We then let the system relax to its steady state.

\begin{figure} [b]
\includegraphics[width=\columnwidth]{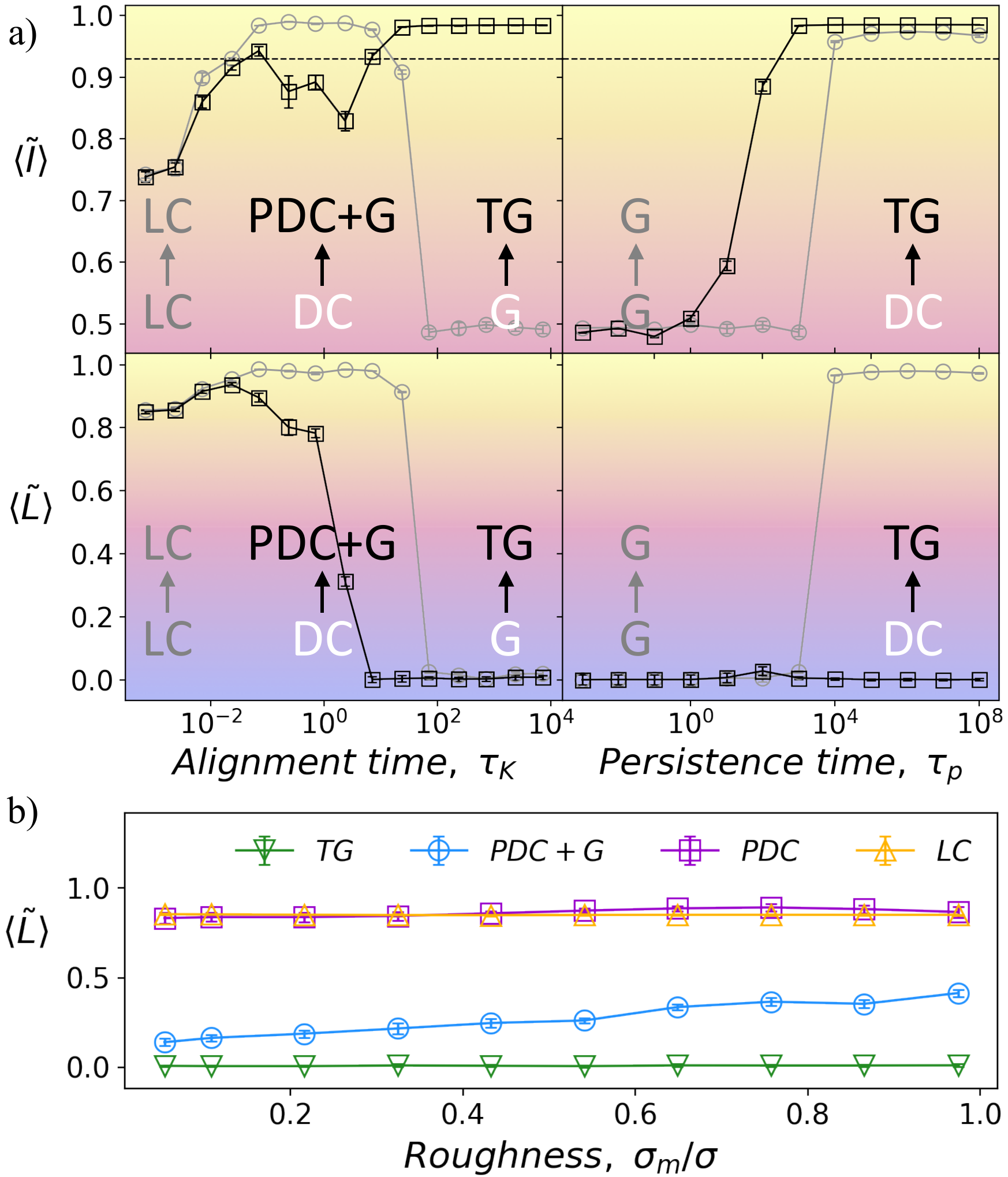}
\caption{(a) Time averages of $\tilde{I}$ and $\tilde{L}$ as functions of $\tau_K$ and $\tau_p$ for smooth (gray circles) and rough (black squares) boundaries. Left panels correspond to $\tau_p = 10^3$ and right panels to $\tau_K = 7.1\times10^1$, representing two cuts of the diagrams in Fig.~\ref{fig:smooth-phases}d and ~\ref{fig:rough-phases}d indicated by the horizontal and vertical lines. The dashed line in the top panels correspond to the value of $\tilde{I}$ for an annulus of thickness $10\sigma$. The background is colored according to the color maps used in Fig.~\ref{fig:smooth-phases}d and \ref{fig:rough-phases}d. (b) Dependence of $\tilde{L}$ on $\sigma_m/\sigma$ for the phases observed in rough boundary conditions at $\tau_p = 10^3$: LC (upward triangles, $\tau_K = 7.1\times10^{-4}$), PDC (squares, $\tau_K = 7.1\times10^{-2}$), coexistence between PDC and gas phases (circles, $\tau_K = 2.4$), and TG (downward triangles, $\tau_K = 7.1\times10^1$).}
\label{fig:moments} 
\end{figure}

Rough boundaries lead to new dynamical phases reported in Fig.~\ref{fig:rough-phases}. Notably, we observe that collisions between particles and boundary monomers favor particles detachment from clusters. While this has little effect on the LC phase, as only a sub-extensive fraction of particles slide along the boundary at any given time, ring-like clusters in the DC phase are strongly disrupted because almost every particle interacts with boundary monomers when roughness is switched on. Consequently, the dependence of the localization parameter $\phi_r^C$ does not show appreciable differences with that observed in smooth boundary conditions. Moving along the vertical line in the diagrams at $\tau_p=10^3$, we observe a partially delocalized clustered (PDC) phase only for $\tau_K = 7.1\times10^{-2}$, where alignment interactions are strong enough to counteract particle detachment from the rotating cluster. Here, the delocalized structure spans roughly $30\%$ of the boundary and is approximately $10\sigma$ thick (see Fig.~\ref{fig:rough-phases}c and Fig.~\ref{fig:moments}a). Note that we observe the PDC phase for $\tau_p < 10^6$, not surviving in the limit of infinite persistence time as the other clustered phases. The rest of the phase space that was once occupied by the DC phase is now divided into two new phases: a coexistence between gas in the bulk and PDC phase along the boundary, and a new phase where particles are stuck near the boundary but the rotation of the center of mass of the system is close to zero. This new trapped gas (TG) phase takes over a portion of the prior gas phase as well. As $\tau_K$ increases and the alignment strength weakens, the system does not recover the homogeneous gas for sufficiently high $\tau_p$. Here, persistence may push particles to reside in the interstitial spaces between boundary monomers long enough to trigger clustering but, since the alignment strength is weak, clusters do not grow in size but instead they continuously form and dissolve. Crucially, $\tilde{I}$ and $\tilde{L}$ are no longer strictly correlated as in the case of smooth boundary conditions, and the TG phase is found for $\tilde{I} \approx 1$ and $\tilde{L} \approx 0$. Thus, these observables are complementary for identifying the TG phase, and only when combined with the localization parameter do they fully capture all the different phases of SPKPs in rough boundary conditions. 

To better quantify the differences between the phases observed for smooth and rough boundaries, we report $\tilde{I}$ and $\tilde{L}$ as functions of $\tau_K$ and $\tau_p$ for both boundary types in Fig.~\ref{fig:moments}a. When roughness is introduced, it is clear that the TG phase takes over a portion of the gas phase where $\tilde{I}$ and $\tilde{L}$ were both nearly zero, see left panels for high $\tau_K$, and a portion of the DC phase where instead $\tilde{I}$ and $\tilde{L}$ were both close to one, see right panels for high $\tau_p$. The transition between the TG and the gas phase is recovered by decreasing persistence time at fixed $\tau_K$ with a much smoother crossover compared to that between the LC and the gas phase. We further examine the dependence of $\tilde{L}$ on boundary roughness by defining the non-dimensional roughness parameter as $\sigma_m / \sigma$, see Fig.~\ref{fig:moments}b. Interestingly, only the coexistence regime between the PDC and the gas phases has an appreciable dependence on roughness. This behavior is due to lower friction between particles and boundary monomers as roughness decreases, which increases the population of particles in the gas that have a contribute negligibly to the angular momentum of the center of mass.

\begin{figure} [t]
\includegraphics[width=\columnwidth]{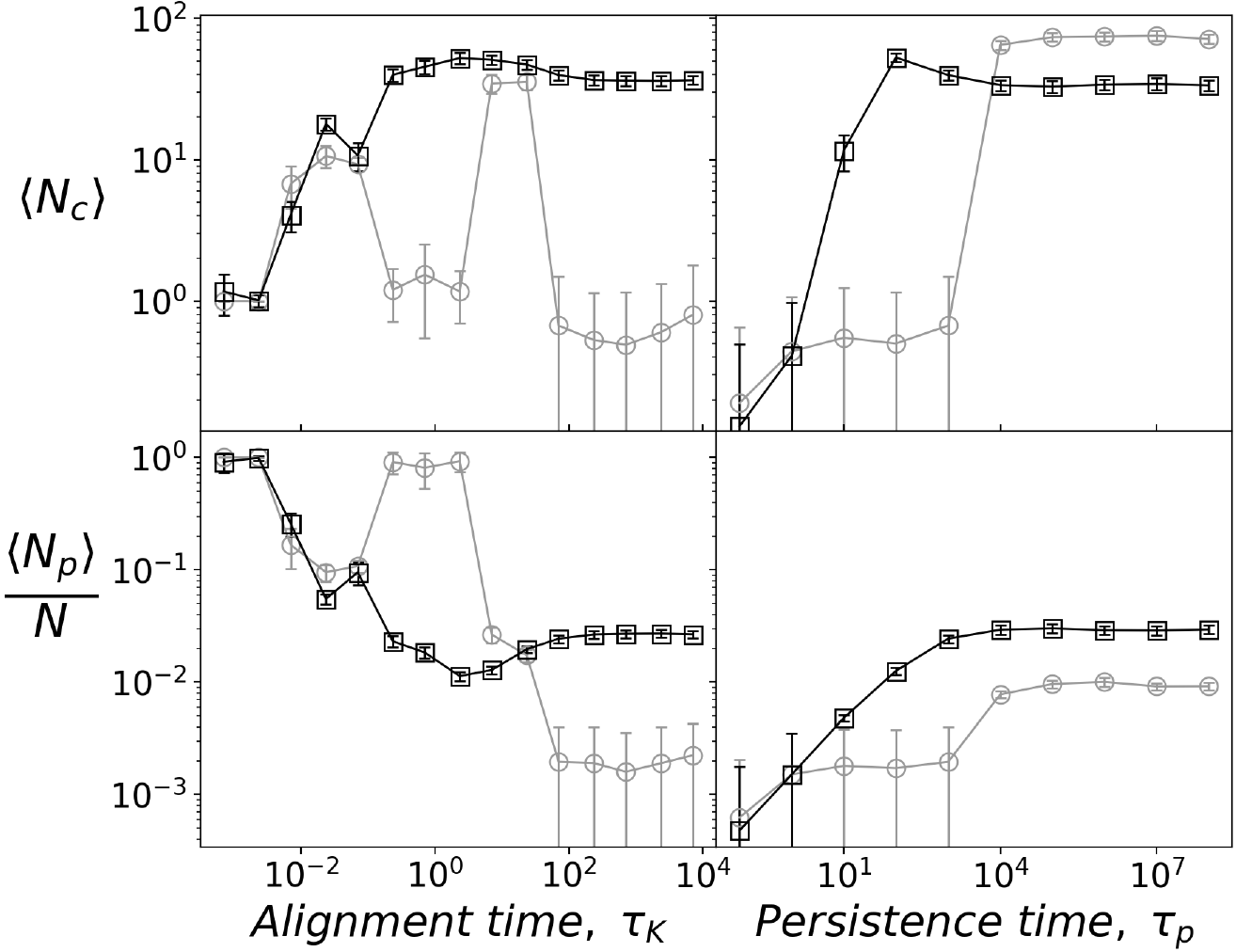}
\caption{Time averages of the number of clusters, $\langle N_c \rangle$, and the cluster fraction, $\langle N_p \rangle/N$, as functions of $\tau_K$ and $\tau_p$ for smooth (gray circles) and rough (black squares) boundaries. Data are obtained from the same simulation runs as those reported in Fig.~\ref{fig:moments}.}
\label{fig:clusters} 
\end{figure}

We conclude our analysis by performing a cluster analysis employing the same method used for computing $\phi_r^C$. Figure~\ref{fig:clusters} shows steady-state averages of the number of clusters, $N_c$, and the average fraction of particles within a cluster, or cluster fraction, $N_p/N$. For both smooth and rough boundaries, the gas phase is identified by negligible values of $N_p/N$. At fixed $\tau_p$ (left panels), and low $\tau_K$, the LC phase, found in both boundary types, corresponds to $N_c \approx N_p/N \approx 1$, meaning that all particles form a single cluster. As alignment strength decreases ($\tau_K$ increases), the number of clusters grows and the cluster fraction decreases, signaling the presence of multiple clusters. When $\tau_K = 7.1 \times 10^{-2}$, clusters start to spread along the wall and boundary friction enters the game. While smooth boundaries allow the formation of a single ring-like cluster where the condition $N_c \approx N_p / N \approx 1$ is met once again, particle-boundary friction leads to coexistence between the PDC and gas phases where $N_c$ is maximum and $N_p/N$ is minimum. When $\tau_K$ is large at fixed $\tau_p$ (left panels) and $\tau_p$ is large at fixed $\tau_K$ (right panels), the system within rough boundaries is in the TG phase, which is characterized by several clusters each containing approximately $2$-$3\%$ of the total number of particles. For smooth boundary conditions instead, when $\tau_p$ is large at fixed $\tau_K$, the system is in the DC phase but the clustering algorithm fails to capture the presence of a single delocalized cluster. More specifically, data reported in the right panels of Fig.~\ref{fig:moments} correspond to relatively weak alignment compared to self-propulsion. This leads to the formation of sparse delocalized clusters where the inter-particle distances are typically larger than those found in compact ring-like clusters for stronger alignment (see Video 5 in~\cite{supplemental}). However, the parameter $\phi_r^C$ remains unaffected by this limitation.

\section{Conclusions}
We have characterized the collective dynamics of SPKPs confined within circular boundaries by systematically exploring the combined effects of self-propulsion and velocity alignment under either smooth or rough boundary conditions. For a purely elastic wall, the interplay between persistence and alignment leads to three distinct dynamical regimes: a gas phase, a delocalized clustered phase characterized by a single ring-like cluster, and a localized clustered phase signaled by the presence of one or more disk-like clusters. Particularly relevant is the strict correlation between the moment of inertia and the angular momentum of the system which are maximized in the delocalized clustered phase. The introduction of boundary roughness fundamentally alters this phenomenology: friction destabilizes delocalized clusters, giving rise to a coexistence regime between multiple clusters on the boundary and the gas phase. Friction also triggers the emergence of a trapped gas phase, where persistent particles form small groups near the boundary that are not able to form a single cluster and thus have negligible center-of-mass angular momentum. These effects can be traced back to momentum exchange between particles and boundary monomers, which disrupts alignment and consequently coherent tangential motion along the confining wall.

Our study proves that boundary friction mediates phases of self-propelled aligning objects. This indicates that the underlying mechanism giving rise to collective motion in confined active systems can be inferred from the macroscopic structures and dynamical patterns that emerge near mechanical obstacles. Delocalized structures signal self-propulsion-dominated motility with weak inter-particle coordination, as observed in \textit{E. coli} performing run-and-tumble motion and self-propelled Janus particles in the dilute regime~\cite{buttinoni_dynamical_2013-2}. By contrast, when self-propulsion is couple with neighbor-to-neighbor alignment, cohesive collective structures emerge—a mechanism whose universality transcends length scales, from synthetic Quincke rollers~\cite{zhang_spontaneous_2023} to large animal groups, e.g. bird flocks, insect swarms and fish schools. In addition, we show that boundary friction can act as a regulatory switch, enabling the system to adapt its accumulation patterns in response to external modifications, such as surface coatings in microfluidic essays of active nematics~\cite{hardouin_reconfigurable_2019}, and rough geometries in experiments of granular active matter~\cite{deseigne_vibrated_2012}. Our results establish a framework for future studies on active systems that are mechanically coupled to their environment, where accumulation near boundaries may trigger rotation as for bacteria in ratchet geometries~\cite{di_leonardo_bacterial_2010} or deformation as in neural crest cells during development~\cite{scarpa_cadherin_2015}.

\section{Acknowledgements}
We thank Flavio Seno for valuable discussions and acknowledge Davide Colì for early contributions.

\section{Data availability}
Data were generated using the in-house developed package \texttt{cudaSoft}, available at \url{https://github.com/farceri/cudaSoft}. The analysis code and data that support the findings of this study are openly available at \url{}.

\bibliography{kuramoto.bib}

\end{document}

\title{Supplemental Material for\\
``Boundary-Mediated Phases of Self-Propelled Kuramoto Particles''}

\author{Francesco Arceri}
\affiliation{INFN, Sezione di Padova, Via Marzolo 8, I-35131 Padova, Italy}
\author{Vittoria Sposini}
\affiliation{Department of Physics and Astronomy, University of Padova, Via Marzolo 8, I-35131 Padova, Italy}
\author{Enzo Orlandini}
\affiliation{Department of Physics and Astronomy, University of Padova, Via Marzolo 8, I-35131 Padova, Italy}
\author{Fulvio Baldovin}
\affiliation{Department of Physics and Astronomy, University of Padova, Via Marzolo 8, I-35131 Padova, Italy}
\date{\today}

\maketitle

% ---------- Numbering with S ----------
\setcounter{figure}{0}
\setcounter{table}{0}
\setcounter{equation}{0}

\renewcommand{\thefigure}{S\arabic{figure}}
\renewcommand{\thetable}{S\arabic{table}}
\renewcommand{\theequation}{S\arabic{equation}}

% ---------- Videos ----------
\section*{Supplemental Videos}

Particles and arrows are color-coded by their velocity orientation according to the rainbow wheel in every video.

\vspace{0.2cm}

\textbf{Video 1.} Phases of SPKPs confined by a smooth boundary: gas phase for $\tau_p = 10^3$ and $\tau_K = 7.1 \times 10^1$, DC phase for $\tau_p = 10^3$ and $\tau_K = 2.4$, and LC phase for $\tau_p = 10^3$ and $\tau_K = 7.1 \times 10^{-4}$. The cluster in the DC phase rotates clockwise while the cluster in the LC phase rotates counter-clockwise. 

\vspace{0.2cm}

\textbf{Video 2.} Phases of SPKPs confined by a rough boundary: gas phase for $\tau_p = 10^{-1}$ and $\tau_K = 7.1 \times 10^1$, TG phase for $\tau_p = 10^3$ and $\tau_K = 7.1 \times 10^1$, coexistence between gas and PDC phase for $\tau_p = 10^3$ and $\tau_K = 2.4$, PDC phase for $\tau_p = 10^3$ and $\tau_K = 7.1 \times 10^{-2}$, and LC phase for $\tau_p = 10^3$ and $\tau_K = 7.1 \times 10{-4}$. The clusters in the PDC and LC phases rotate counter-clockwise. 

\vspace{0.2cm}

\textbf{Video 3.} LC phase for low persistence, $\tau_p = 10^{-1}$, and strong alignment, $\tau_K = 7.1 \times 10^{-4}$ in smooth boundary conditions. The cluster center of mass moves along the average driving direction of the particles that rapidly changes because the persistence time is low. The cluster is not bound to slide along the wall.

\vspace{0.2cm}

\textbf{Video 4.} Multiple clusters in the LC phase for moderate persistence, $\tau_p = 10^3$, and moderately strong alignment, $\tau_K = 7.1 \times 10^{-3}$ in smooth boundary conditions. Depending on their size, clusters merge and dissolve signaling a stable maximum cluster size set by the alignment strength. Playback speed is twice faster than in other videos.

\vspace{0.2cm}

\textbf{Video 5.} DC phase for high persistence, $\tau_p = 10^4$, and weak alignment, $\tau_K = 7.1 \times 10^1$ in smooth boundary conditions. Particles form a sparse delocalized structure due to short alignment time.

% ---------- Additional Figures ----------

\section*{DBSCAN cutoff distance}

The DBSCAN algorithm identifies clusters based on a cutoff distance $r_C$ between neighboring particles and a minimum cluster size. In our analysis we set $r_C = 1.5\sigma$ and require at least 10 particles per cluster. To assess the robustness of the clustering procedure, we tested different values of $r_C$. For configurations characterized by well-defined clusters (Fig.~S1), the results are insensitive to $r_C$ over a wide range, from $1.5\sigma$ to $6\sigma$.

\begin{figure}[h]
\centering
\includegraphics[width=0.9\linewidth]{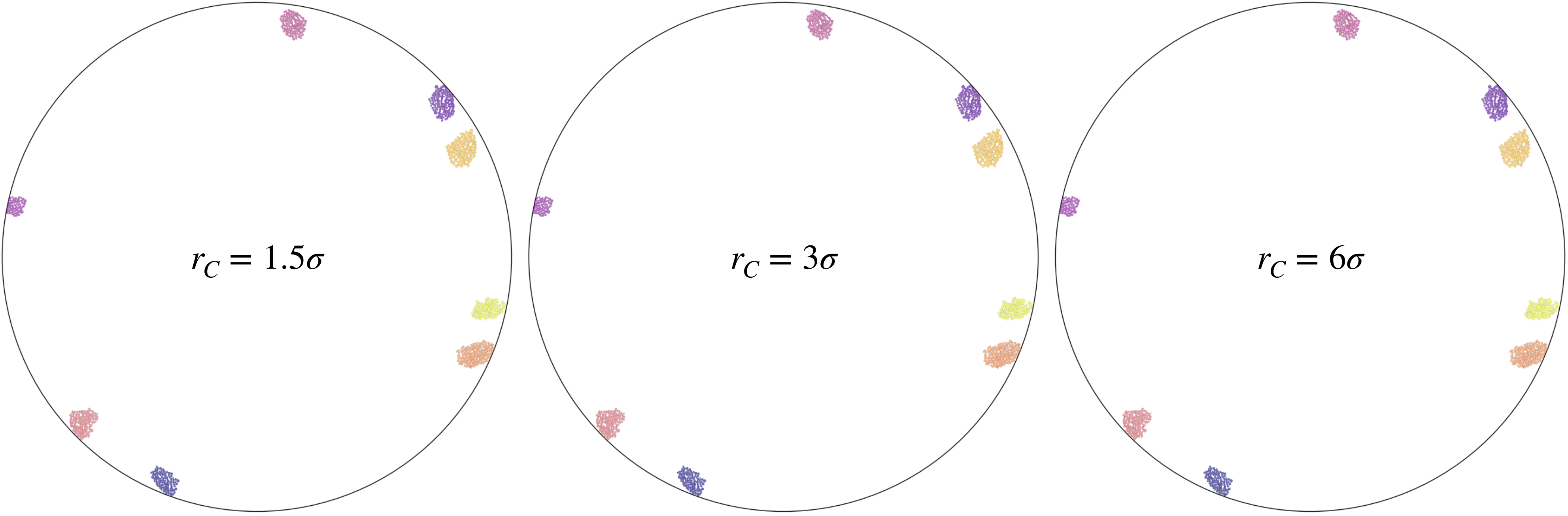}
\caption{Multiple clusters in the LC phase for $\tau_p = 10^3$ and $\tau_K = 2.4 \times 10^{-4}$ in smooth boundary conditions. Particles are color coded by their cluster labels. The identification of the height clusters in the system is insensitive to the choice of $r_C$ for the explored range.}
\end{figure}

By contrast, in configurations where clusters are sparse (Fig.~S2), particles are more weakly connected and the algorithm identifies a single cluster only for larger cutoff distances. 

\begin{figure}[h]
\centering
\includegraphics[width=0.9\linewidth]{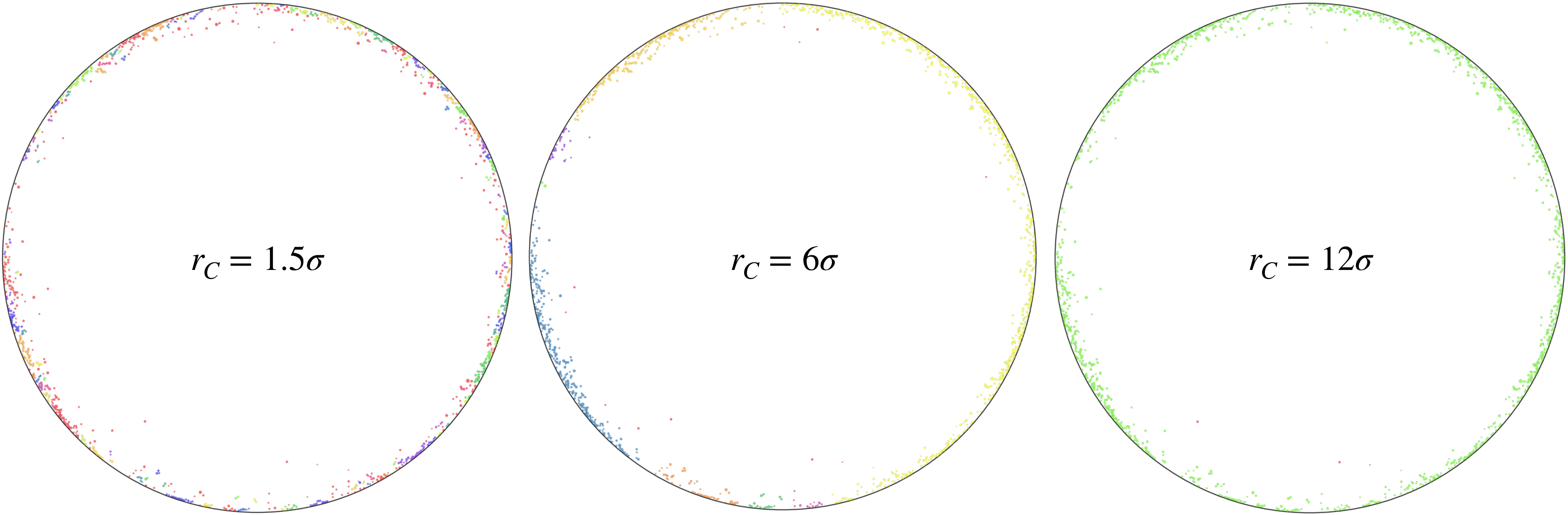}
\caption{A sparse particle configuration in the DC phase for $\tau_p = 10^3$ and $\tau_K = 2.4 \times 10^{-4}$ in smooth boundary conditions. Particles are color coded by their cluster labels. The clustering algorithm identifies a single cluster (rightmost configuration) only for very large cutoff distances.}
\end{figure}

\section*{Smooth boundary with soft interactions}

In the main text we considered two types of boundaries: a smooth, perfectly reflective, circular wall and a rough boundary modeled as a ring polymer whose monomers interact with particles through a WCA potential. We also studied a smooth circular boundary where particles interact with the wall via a WCA potential in each point of the circle of radius $R$, with the same energy scale, $\varepsilon_w = 10\varepsilon$, as that used in rough boundaries. In this case the boundary exerts only normal forces on the particles and therefore introduces no tangential friction, while the interaction remains softer than the perfectly reflective wall. Figure~S3 shows the cluster analysis for this soft boundary. The number of clusters, $N_c$, and the cluster fraction, $N_p/N$, are compatible with those obtained for rough boundaries, indicating that the effects induced by boundary friction smoothly vanish as the tangential forces are removed.

\begin{figure}[h]
\centering
\includegraphics[width=0.6\linewidth]{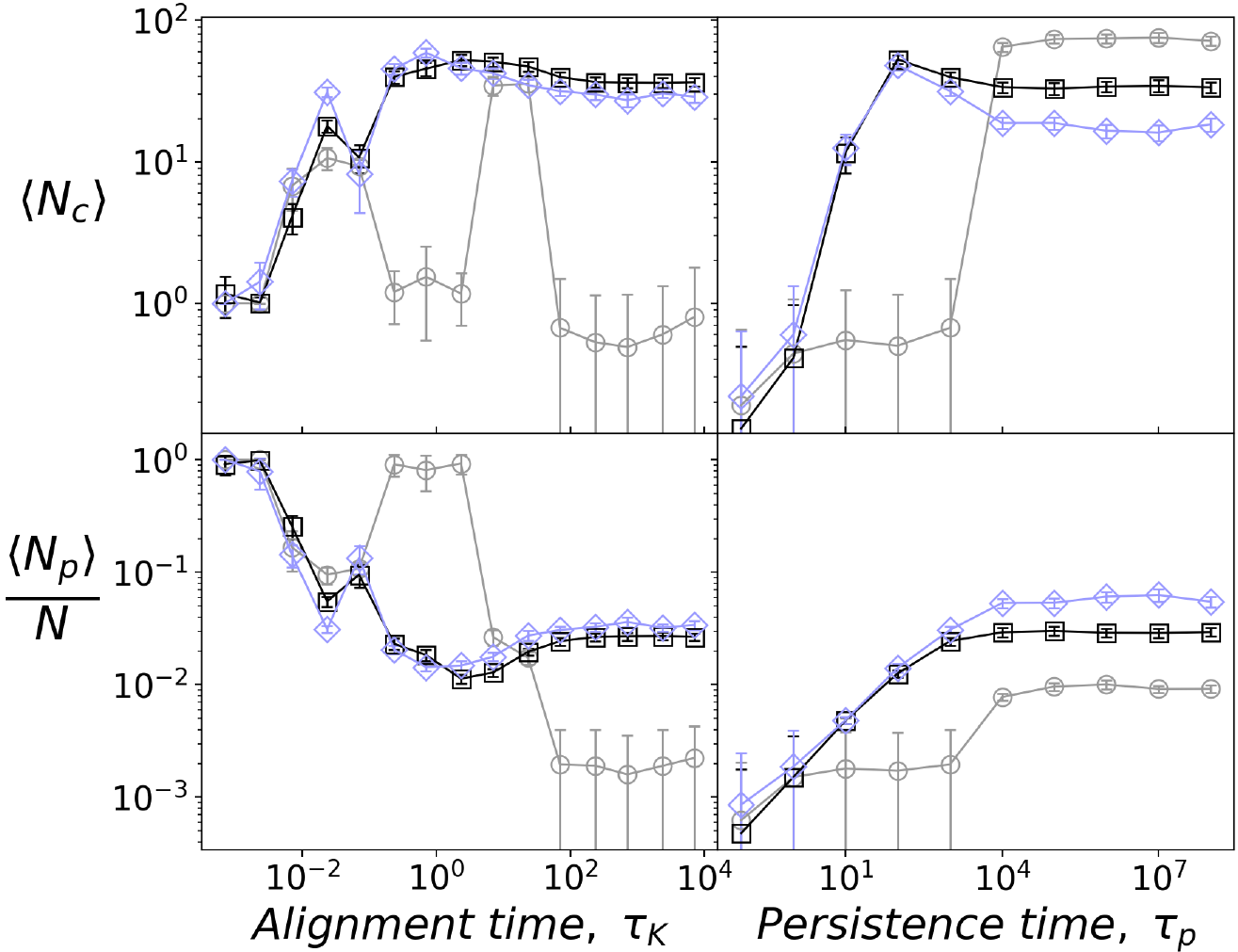}
\caption{Time averages of the number of clusters, $\langle N_c \rangle$, and the cluster fraction, $\langle N_p \rangle/N$, as a function of $\tau_K$ and $\tau_p$ for smooth (gray circles), rough (black squares) and soft (blue diamonds) boundaries.}
\end{figure}